\documentclass[12pt,preprint]{aastex}

\shorttitle{Dust Around WD1150-153}
\shortauthors{Kilic et al.}

\begin{document}

\title{A Dusty Disk Around WD1150$-$153:\\ Explaining the Metals in White Dwarfs by Accretion from the Interstellar Medium versus Debris Disks}

\author{Mukremin Kilic\altaffilmark{1,2,3} and Seth Redfield\altaffilmark{3,4}}

\altaffiltext{1}{Columbus Fellow, Department of Astronomy, Ohio State University, 140 West 18th Avenue, Columbus, OH 43210, USA; kilic@astronomy.ohio-state.edu}

\altaffiltext{2}{Visiting Astronomer at the Infrared Telescope Facility, which is operated by the University of Hawaii under Cooperative Agreement no. NCC 5-538 with the National Aeronautics and Space Administration, Science Mission Directorate, Planetary Astronomy Program.}

\altaffiltext{3}{The University of Texas at Austin, Department of Astronomy, 1 University Station C1400, Austin TX 78712, USA}

\altaffiltext{4}{Hubble Fellow; sredfield@astro.as.utexas.edu}

\begin{abstract}

We report the discovery of excess $K$-band radiation from a metal-rich DAV white dwarf star, WD1150-153. Our near infrared spectroscopic observations
show that the excess radiation cannot be explained by a (sub)stellar companion, and is likely to be caused by a debris disk similar to the other
DAZ white dwarfs with circumstellar debris disks. We find that the fraction of DAZ white dwarfs with detectable debris disks is at least 14\%.
We also revisit the problem of explaining the metals in white dwarf photospheres by accretion from the interstellar medium (ISM). 
We use the observed interstellar column densities toward stars in close
angular proximity and similar distance as DAZ white dwarfs to
constrain the contribution of accretion from the ISM. We find no
correlation between the accretion density required to supply metals
observed in DAZs with the densities observed in their interstellar environment, indicating that ISM
accretion alone cannot explain the presence of metals in nearby DAZ
white dwarfs. Although ISM accretion will certainly contribute, our analysis
indicates that it is not the dominant source of metals for most DAZ white dwarfs.
Instead, the growing number of circumstellar debris
disks around DAZs suggests that circumstellar material may play a more dominant role in polluting the white dwarf atmospheres.

\end{abstract}

\keywords{accretion, accretion disks $-$ circumstellar matter $-$ white dwarfs $-$ stars: individual (WD1150$-$153)}

\section{INTRODUCTION}

White dwarfs can be classified into two main spectral types, DA and DB white dwarfs with hydrogen and helium dominated atmospheres, respectively. DA white dwarfs comprise 86\% of the 9316
spectroscopically confirmed white dwarfs from the Sloan Digital Sky Survey Data Release 4 (SDSS; Eisenstein et al. 2006). The remaining 14\% of the SDSS sample consist of DB,
DQ (1\%), and DZ (1\%) spectral type white dwarfs. The primary physical process affecting the atmospheric composition of white dwarfs is the gravitational separation of elements
in the outer layers of the stars. Due to high surface gravity, gravitational sedimentation is always fast compared to the evolutionary timescales
(Schatzman 1958; Fontaine \& Michaud 1979). Therefore, even if heavy elements are initially present in the atmosphere, they would sink to the bottom of
the photosphere quickly, leaving a pure H or He atmosphere behind.

Radiative levitation and convection can alter the surface atmospheric composition by contaminating the atmosphere with carbon and heavier elements. DQ white dwarfs are the result
of convective dredge up of interior carbon (Weidemann \& Koester 1995). Metals in hot DZ white dwarfs ($T_{\rm eff} >$ 20,000 K) are supported in the atmosphere against gravitational
forces by radiative levitation. However, radiative levitation is expected to be negligible below 20,000 K (Chayer et al. 1995). Yet, we still find white dwarfs with
trace amounts of metals below 20,000 K. In fact, a recent high resolution and high signal-to-noise spectroscopic survey of cool DA white dwarfs by Zuckerman et al. (2003) found
that 25\% of white dwarfs show \ion{Ca}{2} lines. Since the diffusion timescales for heavy elements are short compared to the evolutionary timescales, the observed
metals in cooler DZ white dwarfs cannot be primordial. They must have come from the outside, possibly by accretion from the interstellar medium (Dupuis et al. 1993; Koester \& Wilken 2006, hereafter KW06),
cometary impacts (Alcock et al. 1986), or accretion of asteroidal material from a surrounding debris disk (Graham et al. 1990; Jura 2003).

The ISM is a natural possibility for the source of metals in white
dwarf atmospheres.  It is a continuous and inexhaustible source of
elements, and the range in interstellar volume densities varies over
several orders of magnitude.  For the purposes of ISM accretion onto
white dwarfs to create the DAZ spectral type, understanding the
physical and morphological properties of the local interstellar medium
(LISM), that is material within $\sim$100~pc, is required since almost
all known DAZs reside within this volume. The LISM is an interstellar
environment filled with warm ($T \sim 7000$~K), partially ionized,
moderately dense ($n \sim 0.3$~cm$^{-3}$) material, surrounded by a
volume of hot ($T \sim 10^6$~K), rarefied ($n \sim 0.005$~cm$^{-3}$)
gas known as the Local Bubble. For a review of the properties of the
LISM see Redfield (2006), Ferlet (1999), and Frisch (1995). The LISM
is relatively devoid of cold neutral gas, until the accumulation of
dense material at $\sim$100~pc, roughly the boundary of the Local
Bubble (Lallement et al. 2003).  
However, Meyer et al. (2006) identified one example of a small cold
dense cloud located well within the Local Bubble.

The search for a correlation between LISM clouds and white dwarfs with metals was inconclusive for both DB (Aannestad et al. 1993) and DA white dwarfs (Zuckerman et
al. 2003).
KW06 estimated the ISM densities around known DAZ white dwarfs and suggested that continous accretion
from the warm, partially ionized ISM in the solar neighborhood can explain the accretion of metals onto the DAZ white dwarfs. However, their model requires
a significant variation in the LISM densities and also it cannot explain the metals in DB stars. Observations of helium-rich white dwarfs with metals (DBZ or DZ spectral types)
indicate that in many cases very little or no hydrogen is accreted, suggesting that mostly dust grains are accreted onto the star. Wesemael \& Truran (1982) proposed a propeller
mechanism to prevent accretion of hydrogen onto the DBZ white dwarfs.

An alternative to ISM accretion is accretion of circumstellar disk material, as is the case for G29-38 and GD362. G29-38 and GD362 are DAZ white dwarfs with significant excess infrared
emission from dust disks (Zuckerman \& Becklin 1987; Chary et al. 1999; Reach et al. 2005; Kilic et al. 2005; Becklin et al. 2005; Farihi et al. 2006) that is likely to be the source
of the metals observed in the photosphere of the star.
The case for accretion from debris disks was enhanced by the discovery of the third (GD56, Kilic et al. 2006), and
fourth (WD2115-560, von Hippel et al. 2006) DAZ white dwarf with a circumstellar debris disk (hereafter DAZd; see von Hippel et al. 2006). In addition, Farihi et al. (2006)
found another DAZ (G167-8) with a possible debris disk. 
Including this debris disk candidate, the fraction of known single DAZs with
detectable circumstellar debris disks is 14\%, indicating a possible
source for the observed metals in these white dwarfs. Kilic et
al. (2006) suggested that near- and mid-infrared
observations of more DAZs will increase the fraction with circumstellar disks even further.
Purely based on its temperature and calcium abundance (i.e., its position in their Figure 5), they predicted that WD1150-153, a DAZ star with more metals than
G29-38, is likely to have a $K$-band detectable debris disk.

In this paper, we present our near-infrared spectroscopic observations of WD1150-153 and GD362, and also revisit the problem of accretion scenarios for metals in DAZs.
The spectrum of WD1150-153 and our search for variability in the disk around GD362 are discussed in \S 2. \S 3 describes our efforts to
quantitatively constrain the role of the ISM accretion in DAZ white dwarfs, and results from this analysis are discussed in \S 4.

\section{WD1150-153}

The effective temperatures and calcium to hydrogen abundance ratios for known DAZd white dwarfs (G29-38, GD56, GD362, and WD2115-560) range from 9,700 K to 14,400 K
and 10$^{-7.6}$ to 10$^{-5.2}$, respectively. WD1150-153 has $T_{\rm eff}=$ 12,800 K and (Ca/H) = 10$^{-6.0}$ (KW06). It is the second most
calcium-rich white dwarf after GD362 in the observed DAZd temperature range, and therefore is an excellent
candidate to search for a detectable circumstellar debris disk signature.

WD1150-153 has an apparent $V$ magnitude of 16.0 (Kilkenny et al. 1997) and is estimated to be at a distance of 85 pc (KW06). It
is detected in the Two Micron All Sky Survey (2MASS) in the $J$ and $H$ bands, but not in the $K$ band. It has $J$ = 16.038 $\pm$ 0.119 and $J-H$ = 0.112 $\pm$ 0.210. Due to inaccurate
$J$ and $H$ band photometry and the lack of $K$-band photometry, it is impossible to search for an infrared excess around this object using the 2MASS data alone.

\subsection{Near-Infrared Spectroscopy}

We obtained low resolution near infrared spectra of WD1150-153 over the course of two nights using the 3m NASA Infrared
Telescope Facility (IRTF) equipped with the 0.8--5.4 Micron Medium-Resolution Spectrograph and Imager (SpeX; Rayner et al. 2003).
Our observations were performed under conditions of partly cloudy skies on 22 April 2006 and thin cirrus on 23 April 2006. Our observing and data reduction
procedures are similar to those used by Kilic et al. (2005; 2006).

Figure 1 presents the flux calibrated spectrum of WD1150-153 (second panel) along with the IRTF spectra of the known DAZd white dwarfs with $K$-band excesses.
The observed spectra are shown as black solid lines, and the expected near-infrared photospheric fluxes for each star are shown as red solid lines.
The resolutions of the model white dwarf spectra (kindly provided by D. Koester ) were not matched to the instrumental resolution, in order to show the predicted
locations of the Paschen lines.
Due to non-photometric conditions, the observed $J$ and $H$ band fluxes for WD1150-153 were 17\% lower than the expected fluxes for a 12,800 K white dwarf with $V$ = 16.
We multiplied the IRTF spectrum of this object by 1.17 to match the expected flux level. The non-photometric conditions should not change the overall spectral shape, as Kilic et al. (2006)
did not find any significant effect for the 17 DAZs with apparently normal near-IR spectra obtained under non-photometric conditions. 
The observed white dwarf spectra are affected by strong telluric features between $1.35 - 1.45 \mu$m, $1.80 - 2.05\mu$m, and longward of $2.5 \mu$m.
Paschen lines at 0.955, 1.005, 1.094, 1.282, and 1.876 $\mu$m are apparent in the spectra of DAZs hotter than 10,000 K. 

\subsection{A Dusty Disk Around WD1150-153}

A comparison of the IRTF spectrum of WD1150-153 with the expected flux from a 12,800 K white dwarf demonstrates that the flux
from WD1150-153 is consistent with a typical 12,800 K white dwarf up to 1.7 $\mu$m, but is significantly higher than the expected
flux level from a white dwarf in the $K$-band. The observed excess in WD1150-153 is very similar to the excess seen in known DAZds. 

The expected flux levels from a 12,800 K white dwarf (normalized at $V$ = 16) in the $H$ and $K$ bands are 0.32 and 0.20 mJy, respectively.
The difference between the observed spectrum and the expected flux from the star is 0.10 mJy in the $K$ band. If the excess is caused by a
late type companion, this would correspond to $K \sim$ 17.1 mag. Since the white dwarf is at 85 pc, the absolute $K$-band magnitude of the
companion would be 12.4, an $\sim$L6 dwarf (Leggett et al. 2002). We used an L6 dwarf template from the IRTF Spectral Library
(Cushing et al. 2005) plus the 12,800 K white dwarf model to attempt to match the observed excess in the $K$ band (blue line in the second
panel). Adding an L6 dwarf spectrum to a 12,800 K white dwarf spectrum creates spectral features from 1.3 to 2.4 $\mu$m that are not seen
in the spectrum of WD1150-153, and hence a cool dwarf companion cannot explain the infrared excess.

The IR excess around GD56 and G29-38 can be explained by a $\sim$900 K blackbody (Reach et al. 2005; Kilic et al. 2006). Likewise,
$Spitzer$ observations of GD362 revealed a warm ($\sim$900 K) dust continuum and the likely presence of cooler ($\sim$500 K) dust
(Farihi et al. 2006). The observed excess around WD1150-153 is consistent with a $\sim$900 K blackbody as well. Hence, the most likely
explanation for the $K$ band excess in WD1150-153 is a circumstellar dust disk heated by the white dwarf. We note that G29-38 and GD362
are so far the only DAZs with published Spitzer spectroscopy, which revealed strong silicate emission features and confirmed the presence
of dust disks around these two objects. Kilic et al. (2005) had predicted the existence of a dust disk around GD362 using near-IR
spectroscopic observations. Hence, we are confident that the near-IR excess around WD1150-153 is caused by a dust disk.

We reproduce Figure 5 of Kilic et al. (2006) in Figure 2 with updated calcium abundances and $T_{\rm eff}$ estimates from KW06.
Objects with IRTF near-infrared spectroscopy (filled circles; Kilic et al. 2006 and this study), and Spitzer 4.5 and 8$\mu$m photometry (filled triangles; Mullally
et al. 2006) are also shown. The rest of the DAZs are shown as stars. Objects with circumstellar debris disks (DAZd) are marked with open circles and labeled.
WD1116+026 is also labeled to emphasize the fact that it shows a slight (questionable) $K$-band excess which may be caused by a cooler or nearly edge-on disk (see Figure 2 of
Kilic et al. 2006). Figure 2 demonstrates that 4 of the 5 DAZs with 14,400 K $\geq T_{\rm eff} \geq$ 9,700 K and (Ca/H) $>$ 10$^{-7}$ have debris disks and the remaining object
may be a DAZd awaiting confirmation.

The discovery of a debris disk around WD1150-153 brings the number of DAZds up to 5.
The total number of DAZs observed so far in the near- or mid-infrared is 36. We estimate that the fraction of cool DAZs with detectable debris disks is at least 14\% (5 DAZds
out of 36 white dwarfs). If confirmed, WD1116+026 and G167-8 may bring this fraction up to 19\%.

\subsection{Debris Disk Variability}

The Poynting-Robertson (P-R) effect is expected to cause particles in an optically thin disk to spiral into the white dwarf on timescales of 4$r^2a$ years, where $a$ is
the particle size in microns and $r$ is the distance in solar radii. For G29-38, Reach et al. (2005) estimated that the mid-infrared emission is created by submicron particles
within 1--10 $R_\odot$ of the white dwarf, resulting in a P-R timescale of $\leq$4 years for the inner edge of the disk.
G29-38 is known to have an infrared excess since 1987. Based on the differences between the mid-infrared photometry from the IRTF (Tokunaga et al. 1990), ISO (Chary et al. 1999), and
Spitzer (Reach et al. 2005), von Hippel et al. (2006) point out that the debris disk around G29-38 may have evolved substantially since 1987. However, the IRTF
and ISO data might have been affected by calibration errors. We note that the 2MASS $J, H$, and $K$ band photometry of G29-38 obtained in 2000 is consistent with the photometry of 
Zuckerman \& Becklin (1987) within the errors. 

G29-38 is a ZZ Ceti variable star with up to 27\% amplitude variations in the optical (Bradley 2000). Graham et al. (1990) detected two
significant periodicities in the $K$-band light curve of G29-38 at 181 and 243 seconds with an amplitude of $\sim$0.03 mag. These
periodicities are also seen in the optical light curve of the star and caused by the non-radial g-mode pulsations of the white dwarf.
According to KW06's analysis, WD1150-153 is 700 K warmer than G29-38, which would place it outside of the empirical ZZ Ceti
instability strip for DA white dwarfs determined by Gianninas et al. (2006) and Mukadam et al. (2004). However, recent analyses by Voss (2006) and Gianninas et al. (2006) showed that
its temperature is in the range 12,000-12,450 K. Gianninas et al. (2006) and Koester \& Voss (2006) detected pulsations in the optical light curve of this star with an amplitude
of 0.8\% and a dominant period of 250 seconds. This makes WD1150-153 the second DAV white dwarf with a circumstellar debris disk.
The amplitude variations in the near-infrared are expected to be much less than 0.8\%.


In order to check if the short P-R drag timescales cause any variations in the inner edge of the disk, we compare our near-infrared spectra for WD1150-153 from two consecutive nights, and GD362 spectra
from three different nights in May 2005 and April 2006. Due to non-photometric conditions, our data from 22 April 2006
are less reliable compared to the observations from 23 April 2006. We find that the differences between WD1150-153 spectra from these two nights are 4\% $\pm$ 2\% in the $H$ band
and 6\% $\pm$ 7\% in the $K$ band. The spectra are consistent within the errors and they do not reveal any significant variations.

Figure 3 shows our new spectroscopic data for GD362 compared to that of Kilic et al. (2005). We find that the differences between GD362 spectra from 14 May 2005 and 22 April
2006 are 4\% $\pm$ 1\% in the $H$ band and 9\% $\pm$ 17\% in the $K$ band. On the other hand, GD362 spectrum from 23 April 2006 is more reliable and it only shows a difference
of 1\% $\pm$ 2\% in the $H$ band and 1\% $\pm$ 1\% in the $K$ band compared to the spectrum from 14 May 2005; we did not detect any significant variations in the near-infrared
spectral energy distribution of GD362. 

$K$-band observations of debris disks are most sensitive to the inner, hotter edge of the disk. Therefore, we conclude that the inner edge of the debris
disks around G29-38, GD362, and WD1150-153 seem to be stable (not dynamically, but in terms of physical conditions) on a decadal, yearly, and daily basis, respectively.
If the dust disks around DAZs are optically thick, similar to Saturn's rings, the lifetime of the disks will not be limited to P-R timescales. Von Hippel et al. (2006) suggest
that the lifetime of opaque disks are on the order of 1 Gyr, comparable to the median post-main sequence lifetime of the known DAZds.  

\section{THE ORIGIN OF METAL SPECIES IN DA WHITE DWARFS}

There are two possible accretion sources for the origin of observed metals in white dwarf atmospheres: ISM and circumstellar debris disks.

\subsection{Accretion from the ISM}

Dupuis et al. (1993) proposed a two-phase accretion model in which
white dwarfs accrete at a low rate from warm ISM most of the time, but
high accretion rates take place during brief encounters with
interstellar clouds. The metal abundances decrease exponentially with
the diffusion timescales after the white dwarfs emerge from the ISM
clouds.  This model requires dense neutral clouds within the local
ISM. 
However, few, dense neutral structures are detected in
the LISM (Lallement et al. 2003; Meyer et al. 2006).  KW06 proposed an
alternative scenario for metals; continuous accretion of warm,
partially ionized gas from the ISM.  Assuming that the accretion rates
can be calculated with the Bondi-Hoyle formula, they found the
required ISM densities around DAZs to be in the range $n$(H) = $ 0.01
- 1$ cm$^{-3}$.  Their search for high-metallicity DAZs positioned
close together in the sky turned out to be inconclusive, which lead
them to suggest that the ISM densities must vary by two orders of
magnitude on a scale of a few parsecs to explain the observed
accretion rates onto the DAZs.

Zuckerman et al. (2003) used the \ion{Mg}{2} column densities measured by Redfield \& Linsky (2002) as tracers of the warm gas in the ISM, and searched for a correlation between
the two-dimensional (Galactic latitude and longitude) distribution of DAZs and warm gas on the sky. They did not find any apparent correlation between the lines of sight
with the largest \ion{Mg}{2} column densities and the metal rich DAZs. 

Ideally, one would prefer to have a three dimensional map of the ISM
to make a direct spatial correlation with the observed accretion rates
for white dwarfs. Although our understanding of the three dimensional
structure of the ISM is far from complete, we can still use the
available information to search for a correlation between the warm ISM
and the DAZs. By taking advantage of the growing database of high
resolution observations of LISM absorption toward nearby stars, we are
able to determine empirically the interstellar environment (e.g.,
volume density) around each known DAZ white dwarf.  Ultraviolet (UV)
absorption lines are very sensitive to even the low column density
structures common in the LISM (Redfield 2006).  We have searched
through the high resolution LISM UV absorption line database (e.g.,
Redfield \& Linsky 2002; Redfield \& Linsky 2004) to identify stars
that are close to known DAZ white dwarfs in angle (2.1$^{\circ} <
\Delta\theta < 29.1^{\circ}$; median $\Delta\theta = 16.9^{\circ}$)
and distance (0.5 pc $< \Delta d <$ 73.3 pc; median $\Delta d =
8$~pc).  When possible, we have tried to bracket the white dwarf
distance in order to make a more reliable estimate of the volume
density in the vicinity of the DAZ white dwarf.  Future high
resolution observations are needed of the LISM in the vicinity of
these objects in order to better constrain the properties of the
interstellar material these WDs have recently encountered.

Table 1 presents distances, Galactic coordinates, and the observed
volume densities toward DAZ white dwarfs and their nearby reference
stars.  The theoretically predicted volume densities required to
account for the metals in white dwarfs from accretion of ISM material
(KW06; kindly provided by D. Koester) are given as $\log n$(H)$_{\rm
pred}$.  $\Delta\theta$ is the angle between the white dwarf and the
reference star, $\log N$(H){\Large$_\star$} and $\log
n$(H){\Large$_\star$} are the column and volume densities observed
toward the reference stars, respectively.  Typically, $N(\rm H)$ is
not measured directly, but is estimated from more easily observed ions
(e.g., \ion{D}{1}, \ion{Mg}{2}, \ion{Ca}{2}, \ion{Fe}{2} etc; Redfield
\& Linsky 2002, 2004; Frisch, Grodnicki, \& Welty 2002).  We take the
LISM average abundance ratios ($N(X)/N(\rm H)$) to convert these ions
into hydrogen column densities.  The LISM D/H abundance ratio is
remarkably constant (Moos et al. 2002; Linsky et al. 2006), and
therefore we try to use \ion{D}{1} when possible to estimate $N(\rm
H)$. 59\% of our sample has \ion{D}{1} measurements, where the
root-mean-square ({\emph RMS}) is 10\% of the median abundance ratio,
and therefore is a negligible contribution to the total error budget
of our density calculations. However, when \ion{D}{1} observations do
not exist, we are forced to use other observed ions, where the
abundance ratio relative to hydrogen varies more dramatically in the
LISM.  We use \ion{Ca}{2} as a proxy in 32\% of our measurements, were
the {\emph RMS} is 120\% of the median abundance ratio, \ion{Fe}{2} in
6\% of our measurements, where the {\emph RMS} is 90\% of the median
abundance ratio, and \ion{Mg}{2} in 3\% of our measurements, where the
{\emph RMS} is 230\% of the median abundance ratio.  We incorporate
the {\emph RMS} of the abundance variation as an estimate of the
systematic error in our conversion to hydrogen densities.  Although it
is typically the dominant source of error if we are using \ion{Mg}{2},
\ion{Ca}{2}, or \ion{Fe}{2} as a proxy for hydrogen, it rarely exceeds
the range of densities calculated from a pair of reference stars.  The
average volume densities for reference stars are obtained simply by
dividing the column densities with distances to the stars. We omitted
three white dwarfs from the DAZ list in KW06 due to lack of distance
estimates.  In addition, six white dwarfs lacked U, V, and W space
velocity measurements, which are required for estimating the mass
accretion rate onto the white dwarfs using the Bondi-Hoyle
formulation. This limited the accuracy of $\log n$(H)$_{\rm pred}$
estimates for these six white dwarfs.  Therefore, we limit our
analysis to the remaining 28 DAZs.

Figure 4 shows the theoretically predicted volume density for these 28 DAZ white dwarfs versus the observed average
volume densities for warm, partially ionized ISM measured from the nearby reference stars.
Objects hotter than 11,000 K are shown as filled circles and the remaining objects are shown as star symbols.
For objects with two or more reference stars, we have linearly interpolated the column densities
for the reference stars to estimate the column densities at white dwarfs distance.
The ``error bars'' include the propagated errors of column density, distance, and abundance ratio to hydrogen and they
show the observed range of ISM volume densities for the reference stars for each white dwarf.

Figure 4 shows that there is no apparent correlation between the
theoretically required ISM densities around DAZs and the average
volume densities observed towards stars near DAZs. 
This result is significant for the hottest objects with
$T_{\rm eff}>$ 11,000 K for which the diffusion timescales are less than 6 months. In fact, the diffusion timescales for the majority of these objects
are on the order of 10 days. If the accretion from the ISM is responsible for the metals in these stars,
we would expect them to be embedded in at least modestly dense ISM.  The
ratio of the theoretically predicted volume density to the observed
average density ranges from 0.02 to 428 for these DAZs.
However, the average volume densities assume a constant density over path lengths
that are much larger than the expected size of LISM structures
(Redfield \& Linsky 2000).  In addition, the observed path length extends
from the DAZ to the Earth, and may be dominated by material along the
sightline far from the vicinity of the DAZ.  Ideally, we would like to
measure the volume density right at the location of the white dwarf.
Therefore, we have identified a subset of the DAZs that have reference
stars that bracket the white dwarfs, in order to more accurately
determine the volume density in the vicinity of the DAZ.  It is
important to remember that these are lower limits to the volume
density, since it is unlikely that the ISM density is constant over
even these path lengths. 
WD0245+541 might be a good example, where the column density increases from the reference star closer than the white dwarf ($\eta$ Cas A) to the star more distant than
the white dwarf (HR 1925), and therefore given the difference in the distances and the change in column densities, we can calculate the average volume density in the
vicinity of the white dwarf. However, this can only be done for a subset of our targets. For some of our targets, due to small scale morphological variations, more
distant reference stars have lower column densities compared to the nearby reference stars.

Figure 5 shows the average ISM volume densities observed in the
vicinity of 14 DAZ white dwarfs\footnote{G167-8 (=WD1455+298), Farihi et al.'s debris disk candidate, has an order of magnitude more metals than its
surrounding ISM, which favors the presence of a debris disk.}. The small number statistics and large error bars prevent
us from a definitive conclusion, however this figure confirms the
analysis from Figure 4 in the sense that we do not see any apparent
correlation between the theoretically predicted ISM densities
required to account for metals in DAZs and the average densities
observed in the gas surrounding DAZs in the LISM.
In addition, WD0245+541 and WD1202-232 present two cases
where the current density of interstellar material in the vicinity of
these white dwarfs ($\log n$(H)$_{\rm ISM}= -1.28^{+0.20}_{-0.38}$ and $-1.17^{+0.03}_{-0.04}$) greatly exceeds
that predicted to account for the metals in their atmospheres ($\log n$(H)$_{\rm pred}= -3.00$ and $-2.18$).
The diffusion timescales for these two stars are $\sim$30,000 yr and 400 yr, respectively. We would expect them to have
at least the same amount of metals as their surrounding ISM.
The difference could be caused if ISM accretion occurs differently than assumed
by KW06, for example, astrospheric structures (Wood 2004) around the
white dwarf may filter or accumulate material, changing the rate and
properties of accretion.

There is evidently not the appropriate distribution of LISM material
to account for the metals in DAZ white dwarfs solely from ISM
accretion.  We note that the LISM is likely patchy, and therefore our
average volume densities may be severe underestimates of the true
densities in the vicinity of the DAZs. For example, if a small dense
cloud, similar to one recently uncovered by Meyer et al. (2006) is
located in the vicinity of a WD, it may go completely undetected, or
the density underestimated because we are unable to tightly constrain
the physical size of the cloud. 
On the other hand, since DA and DAZ white
dwarfs share the same volume and similar dynamical properties
(Zuckerman et al. 2003), a purely spatial solution, such as accretion
of interstellar material, seems unlikely to be a comprehensive
explanation for metals in DAZ white dwarf atmospheres. However,
densely sampled surveys of LISM material are required to adequately
reconstruct the interstellar environments in the vicinity of nearby
WDs.  Although accretion of material from the surrounding ISM will
certainly contribute, our analysis using the currently most complete
LISM surveys, indicate that it is not the dominant source of metals
for at least most DAZ white dwarfs.

\subsection{Accretion from Circumstellar Debris Disks}

The discovery of several circumstellar debris disks around metal-rich white dwarfs and the non-detection of debris disks around other white dwarfs (Mullally et al.
2006) suggest a strong connection between the ``Z'' phenomenon and debris disks. However, one question remains to be answered: Is there enough material in these disks to explain
the observed accretion rates of metals onto the white dwarfs?

A detailed analysis of the mid-infrared spectral energy distribution of G29-38 (Reach et al. 2005) showed that the total dust
mass is on the order of 10$^{18}$ g (4\%--50\% of the zodiacal cloud; Fixsen \& Dwek 2002). The dust mass can be higher if there are larger, cooler grains that do
not contribute to the observed flux in the mid-infrared. KW06 estimate an accretion rate of 10$^{-15.13} M_\odot$ yr$^{-1}$ assuming that the accreted
material has solar composition. If the accreted material is mostly refractory material, the accretion rate would be $\sim$ 50 times smaller, corresponding to
2.9 $\times$ 10$^{16}$ g yr$^{-1}$. There is enough material in the disk around G29-38 to supply metals to the star for 34 years; the small grains that make
up most of the mid-infrared emission have to be continuously regenerated by collisions of larger bodies. 
If we assume that G29-38 has been accreting at the same rate since it has evolved into a white dwarf 500 million years ago, the total disk mass required would be
2 $\times$ 10$^{-3} M_\earth$, similar to the mass of Pluto. The total disk mass required for the other DAZd white dwarfs, except GD362, range from 0.3 to
2.3 Pluto masses. GD362 is a $\geq$ 2.5 Gyr old massive white dwarf with nearly solar metal abundance. Assuming continous accretion of metals for its entire lifetime
would require a Uranus-mass, terrestrial type planet to be accreted by the white dwarf.

\section{DISCUSSION}

Our analysis shows that there is no apparent correlation between the local ISM densities
surrounding cool DAZ white dwarfs, and the metals in their atmospheres.
In addition, Jura (2006) argues that there are three DZ white dwarfs with carbon to iron abundance ratios at least an order of magnitude below solar,
which cannot be explained by accretion from the ISM. The carbon-deficient composition is easily explained if these stars accreted
asteroids with a chondritic composition.

We expect most of the planets in our solar system to survive the red giant phase of the Sun. Mercury and Venus are most likely to be engulfed in the envelope of the red giant Sun.
Earth's orbit would expand due to stellar mass loss, however tidal interactions between the earth and the enlarged Sun may cause the Earth to spiral into the Sun (Rasio et al. 1996).
Using detailed simulations of the evolution of planetary orbits through the solar mass loss phase, Duncan \& Lissauer (1998) showed that the orbits of the
terrestrial planets that survive the Sun's red giant phase are stable for more than 1 Gyr assuming plausible solar mass loss. Likewise, many of the larger asteroids that survive
the AGB phase seem to be stable, although some of the asteroids, e.g. Pallas, will become unstable within 500 million years, showing secular growth in eccentricity. 
If the progenitors of white dwarfs had asteroid belts similar to the one in our solar system, and if these asteroids survived the AGB phase, we would expect some of them to
be dynamically unstable. An asteroid's initial relatively circular orbit may be altered after the AGB phase with the consequence that it may end up within the Roche radius of the
white dwarf to be tidally disrupted to create a debris disk (Jura 2003). An asteroid like Pallas would supply the white dwarf with 2.2 $\times$ 10$^{23}$ g of refractory material, which
would be enough to supply metals to G29-38 for 10$^7$ years. 

Duncan \& Lissauer (1998) argued that massive stars are expected to lose a larger fraction of their mass during post-main sequence evolution, and planetary
systems around massive stars may be destabilized due to the mass loss process. The progenitors of the known DAZd white dwarfs were more massive than the Sun.
If these progenitor stars had planetary systems that survived the AGB phase, these systems might have been destabilized due to mass loss.
Debes \& Sigurdsson (2002) suggested that the planets around white dwarfs may become unstable to close approaches with each other and the entire system may become
dynamically young. They predict that the orbits of comets, asteroids, and Kuiper Belt objects (KBOs) will be perturbed and the white dwarf would experience a late bombardment episode.
The white dwarf progenitor's high luminosity during its red giant phase is likely to vaporize a large fraction of the icy KBOs.
However, the rocky cores of such objects might survive and eventually end up
close to the white dwarf (Jura 2006). Debes \& Sigurdsson (2002) predicted that the heavy bombardment by comets/KBOs starts 10--100 million years after the mass loss phase and
gradually declines on timescales of 0.1--1 billion years. They expect 0.1\% of the KBOs to enter the inner system. If we assume that the white dwarfs have
Kuiper Belts similar to our own, we would expect 8 $\times$ 10$^{-5} M_\earth$ to 1 $\times$ 10$^{-3} M_\earth$ (Luu \& Jewitt 2002) of material to end up in the inner
system and eventually accreted by the white dwarf. Accretion of the refractory material from these objects would be enough to feed G29-38 for 10$^7$--10$^8$ years.
Since the progenitors of the known DAZd white dwarfs were more massive than the Sun, a more massive Kuiper Belt or cometary system is possible for these white dwarfs. A
massive Kuiper Belt would also provide enough mass for the disk lifetime to be consistent with the ages of the DAZd white dwarfs.
 
Another possible scenario for explaining dust around white dwarfs is a fallback disk from the AGB phase. The silicate feature
observed in the spectral energy distribution of G29-38 is similar to the solar system zodiacal light and the oxygen-rich
mass loss from the star Mira. Since the disk around G29-38 is within
1--10 $R_\odot$, if the source of the debris is leftover material from the AGB phase, it must have been transported inward. 
A debris disk is the presumed origin of the planets detected around the millisecond pulsar PSR B1257+12 (Wolszczan 1994).
In addition, a fallback disk is observed around a young neutron star (Wang et al. 2006). A similar mechanism would explain debris disks around
hot, young white dwarfs. However, none of the hot white dwarfs in recent Spitzer surveys of Hansen et al. (2006) and Mullally et al. (2006)
have detectable debris disks. In addition, all known DAZd white dwarfs are older than 200 million years old, and GD362 is estimated
to be older than 2.5 billion years. Therefore, the best explanation for dust disks seem to be tidal disruption of comets, asteroids,
or minor planets that are perturbed into high eccentricity orbits by the remnant planetary systems of white dwarfs.

\section{CONCLUSIONS}

Metals are observed in the atmospheres of several cool DA white
dwarfs.  Because diffusion timescales are much shorter than the
evolutionary lifetimes of these objects, a constant source of metals
is required.  In order to assess the influence of accretion from the
local ISM, we estimated the local volume density in the vicinity of
several DAZ stars.  We find no correlation between the accretion
density required to supply metals observed in DAZ white dwarfs with
the densities observed in their interstellar environment, indicating
that ISM accretion alone cannot explain the presence of metals in
their atmospheres.  Despite likely ISM accretion, it appears that this
is not the dominant source of metals for most DAZ white dwarfs.

Recent discoveries of 5 DAZd white dwarfs, including WD1150-153, shows that accretion from circumstellar debris disks can explain the metals in at least 14\% of the DAZs.
We also find evidence for high accretion rates onto G167-8 (compared to its surrounding ISM), which suggests that the observed mid-IR excess around this white dwarf (Farihi et al. 2006)
may be caused by a debris disk.
Accretion of asteroids and minor planets like Pluto would provide enough material to explain the observed accretion rates onto the DAZ white dwarfs.
The fraction of sun-like stars possessing planets is $\geq$5\%, and it can be as much as 100\% (Lineweaver \& Grether 2003). 
DAZ white dwarfs make up 25\% of all DA white dwarfs. The observed fraction of DAZds ($\geq$14\%) among DAZ white dwarfs suggests that $\geq$3.5\% of DA white dwarfs
may have remnant planetary systems.
Mid-infrared observations of more DAZs may end up finding debris disks around other cool DAZs, offering at least a partial solution for the origin of metals in white dwarf photospheres.

\acknowledgements
We would like to thank Ted von Hippel and Andrew Gould for helpful discussions. We also thank our anonymous referee for useful suggestions.
S.R. would like to acknowledge support provided by NASA through Hubble Fellowship grant HST-HF-01190.01 awarded by the
Space Telescope Science Institute, which is operated by the Association of Universities for Research in Astronomy, Inc., for NASA, under contract NAS 5-26555.

\begin{deluxetable}{lcrrccrrrrcc}
\tabletypesize{\small}
\rotate
\tablecolumns{12}
\tablewidth{0pt}
\tablecaption{DAZ White Dwarfs and Nearby Stars}
\tablehead{
\colhead{White Dwarf}&
\colhead{$d_{\rm WD}$}&
\colhead{\it l$_{\rm WD}$}&
\colhead{\it b$_{\rm WD}$}&
\colhead{$\log n$(H)$_{\rm pred}$}&
\colhead{Nearby Star}&
\colhead{\it l\Large$_\star$}&
\colhead{\it b\Large$_\star$}&
\colhead{$d$\Large$_\star$}&
\colhead{$\Delta\theta$}&
\colhead{$\log N$(H)\Large$_\star$}&
\colhead{$\log n$(H)\Large$_\star$}
\\
 & (pc) & & &(cm$^{-3}$) & & & & (pc) & (deg) & (cm$^{-2}$) & (cm$^{-3}$)
}
\startdata
WD0032$-$175 & 32 & 101.3 & $-$79.5 & $-$2.162 & $\beta$ Cet & 111.3 & $-$80.7 & 29.4 & 2.09 & 18.52 & $-$1.44 $^{+0.16}_{-0.32}$\\
HE0106$-$3253 & 66 & 270.2 & $-$83.4 & $-$0.501 & $\alpha$ Eri & 290.8 & $-$58.8 & 44.1 & 25.12 & 18.43 & $-$1.71 $^{+0.32}_{-0.68}$ \\
WD0235+064 & 34 & 164.4 & $-$47.5 & $-$4.389:  & HR 1099 & 184.9 & $-$41.6 & 29.0 & 15.72 & 18.10 & $-$1.85 $^{+0.03}_{-0.04}$ \\
           &    &        &        & & V471 Tau & 172.5 & $-$27.9 & 46.8 & 20.56 & 18.21 & $-$1.95 $^{+0.04}_{-0.03}$ \\
WD0243$-$026 & 21 & 176.0 & $-$53.0 & $-$1.298   & EP Eri & 192.1 & $-$58.3 & 10.4 & 10.45 & 17.95 & $-$1.56 $^{+0.09}_{-0.11}$ \\
           &    &        &        & & HR 1099 & 184.9 & $-$41.6 & 29.0 & 12.90 &  18.10 & $-$1.85 $^{+0.03}_{-0.04}$ \\
WD0245+541 & 10 & 139.5 & $-$4.7  & $-$3.000      & $\eta$ Cas A & 122.6 & $-$5.1 & 6.0 & 16.88 & 17.90 & $-$1.37 $^{+0.24}_{-0.60}$ \\
           &    &        &        & & HR 1925 & 158.4 & +12.0 & 12.2 & 25.08 & 18.26 & $-$1.32 $^{+0.01}_{-0.01}$ \\
WD0408$-$041 & 74 & 196.1 & $-$37.1 & $-$0.346     & HR 1608 & 209.6 & $-$29.4 & 54.7 & 13.67 & 18.07 & $-$2.15 $^{+0.05}_{-0.05}$ \\
WD0543+579 & 30 & 154.8 & +14.9   &  $-$1.160:     & $\beta$ Aur & 167.5 & +10.4 & 25.2 & 13.21 & 17.64 & $-$2.25 $^{+0.32}_{-0.66}$ \\
           &    &        &        & & DK UMa & 142.6 & +38.9 &  32.4 & 26.26 &  17.95 & $-$2.05 $^{+0.02}_{-0.02}$ \\
WD0846+346 & 27 & 189.0 & +38.3   &  $-$2.232:      & $\sigma$ Gem & 191.2 &  +23.3 &  37.5 & 15.13 &  18.12 & $-$1.94 $^{+0.12}_{-0.16}$ \\
           &    &        &        & & HR 4345 &  184.3 &  +67.3 &  21.7 & 29.11 &  17.82 & $-$2.01 $^{+0.02}_{-0.02}$ \\
WD1015+161 & 95 & 222.5 & +52.8   &  +0.621      & HR 4345 &  184.3 &  +67.3 &  21.7 & 23.37 &  17.82 & $-$2.01 $^{+0.02}_{-0.02}$ \\
WD1102$-$183 & 40 & 270.5 & +37.5 &  $-$2.215      & $\gamma$ Crv & 291.0 &  +44.5 &  50.6 & 16.92 &  18.48 & $-$1.72 $^{+0.32}_{-0.71}$ \\
           &    &        &        & & HR 4657 &  288.5 &  +51.5 &  22.6 & 18.91 &  18.57 & $-$1.27 $^{+0.02}_{-0.03}$ \\
WD1116+026 & 43 & 257.3 & +56.8   &  $-$0.871       & HR 4657 &  288.5 &  +51.5 &  22.6 & 18.84 &  18.57 & $-$1.27 $^{+0.02}_{-0.03}$ \\
           &    &        &        & & $\gamma$ Crv & 291.0 &  +44.5 &  50.6 & 24.30 &  18.48 & $-$1.72 $^{+0.32}_{-0.71}$ \\
WD1124$-$293 & 34 & 281.9 & +29.7 &  $-$0.267       & $\mu$ Vel &  283.0 &   +8.6 &  35.3 & 21.16 &  18.47 & $-$1.57 $^{+0.01}_{-0.01}$ \\
           &    &        &        & & HR 4657 &  288.5 &  +51.5 &  22.6 & 22.33 &  18.57 & $-$1.27 $^{+0.02}_{-0.03}$ \\
WD1150$-$153 & 85 & 282.9 & +45.0 &  $-$1.197        & $\alpha$ Vir & 316.1 &  +50.8 &  80.4 & 22.78 &  18.40 & $-$2.00 $^{+0.32}_{-0.68}$ \\
WD1202$-$232 & 10 & 289.5 & +38.1 &  $-$2.183        & HR 4657 &  288.5 &  +51.5 &  22.6 & 13.40 &  18.57 & $-$1.27 $^{+0.02}_{-0.03}$ \\
           &    &        &        & & 61 Vir &  311.9 &  +44.1 &   8.5 & 17.84 &  17.91 & $-$1.51 $^{+0.04}_{-0.04}$ \\
WD1204$-$136 & 52 &  286.8 & +47.6 & $-$0.141        & HR 4657 &  288.5 &  +51.5 &  22.6 &  4.07 &  18.57 & $-$1.27 $^{+0.02}_{-0.03}$ \\
           &    &        &        & & $\gamma$ Crv &  291.0 &  +44.5 &  50.6 &  4.24 &  18.48 & $-$1.72 $^{+0.32}_{-0.71}$ \\
           &    &        &        & & $\alpha$ Vir & 316.1 &  +50.8 &  80.4 & 19.28 &  18.40 & $-$2.00 $^{+0.32}_{-0.68}$ \\
WD1208+576 & 20 & 133.4 & +58.9   &  $-$1.628        & SAO 28753 & 113.7 &  +59.5 &  21.9 & 10.04 &  18.19 & $-$1.64 $^{+0.02}_{-0.02}$ \\
           &    &        &        & & CF UMa &  168.5 &  +73.8  &  9.2 & 19.92 &  17.70 & $-$1.75 $^{+0.37}_{-0.93}$ \\
HE1225+0038 & 29 & 290.2 & +62.7  &  $-$2.293        & HR 4657 &  288.5 &  +51.5 &  22.6 & 11.20 &  18.57 & $-$1.27 $^{+0.02}_{-0.03}$ \\
           &    &        &        & & HR 4803  & 299.2 &  +35.6 &  34.6 & 27.64 &  18.96 & $-$1.06 $^{+0.30}_{-0.86}$ \\
WD1257+278 & 35 &  46.7 & +88.1   &  +0.017        & HZ 43  &  54.1 &  +84.2 &  32.0 &  3.91 &  17.96 & $-$2.04 $^{+0.14}_{-0.11}$ \\
           &    &        &        & & GD 153 &  317.3 &  +84.8 &  70.5 &  5.52 &  17.94 & $-$2.40 $^{+0.03}_{-0.04}$ \\
HE1315$-$1105 & 40 & 313.2 & +51.0 & $-$3.004        & $\gamma$ Crv & 291.0 &  +44.5 &  50.6 & 16.21 &  18.48 & $-$1.72 $^{+0.32}_{-0.71}$ \\
           &    &        &        & & SAO 158720 & 337.5 &  +39.2 &  23.6 & 20.64 &  18.11 & $-$1.75 $^{+0.02}_{-0.03}$ \\
WD1337+705 & 33 & 117.2 & +46.3   &  +0.007        & SAO 28753 & 113.7 &  +59.5 &  21.9 & 13.35 &  18.19 & $-$1.64 $^{+0.02}_{-0.02}$ \\
           &    &        &        & & DK UMa &  142.6 &  +38.9 &  32.4 & 20.01 &  17.95 & $-$2.05 $^{+0.02}_{-0.02}$ \\
WD1344+106 & 18 & 343.8 & +68.7   &  $-$2.257        & $\xi$ Boo A  &  23.1 &  +61.4 &   6.7 & 17.73 &  17.87 & $-$1.45 $^{+0.03}_{-0.03}$ \\
           &    &        &        & & HZ 43  &  54.1 &  +84.2 &  32.0 & 20.10 &  17.96 & $-$2.04 $^{+0.14}_{-0.11}$ \\
WD1407+425 & 26 &  81.5 & +68.0   &  $-$3.470:        & $\eta$ UMa &  100.7 &  +65.3 &  30.9 &  8.05 &  17.91 & $-$2.07 $^{+0.03}_{-0.04}$ \\
           &    &        &        & & SAO 28753 & 113.7 &  +59.5 &  21.9 & 16.32 &  18.19 & $-$1.64 $^{+0.02}_{-0.02}$ \\
WD1455+298 & 35 &  45.6 & +62.1   &  +0.079        & $\beta$ Ser  &  26.0 &  +47.9 &  46.9 & 17.97 &  18.66 & $-$1.50 $^{+0.32}_{-0.66}$ \\
           &    &        &        & &  HZ 43  &  54.1 &  +84.2 &  32.0 & 22.17 &  17.96 & $-$2.04 $^{+0.14}_{-0.11}$ \\
WD1457$-$086 & 117 & 348.3 & +42.5 &  $-$0.479        & HD141569 &  4.2 &  +36.9 &  99.0 & 13.40 &  19.29 & $-$1.19 $^{+0.32}_{-0.73}$ \\
WD1614+160 & 117 &  30.7 & +41.2  &   $-$0.764        & HD141569 &    4.2 &  +36.9 &  99.0 & 20.93 &  19.29 & $-$1.19 $^{+0.32}_{-0.73}$ \\
WD1633+433 & 15  &  68.1 & +42.5  &   $-$0.756        & $\chi$ Her &   67.7 &  +50.3 &  15.9 &  7.78 &  18.25 & $-$1.44 $^{+0.01}_{-0.01}$ \\
           &    &        &        & & $\gamma$ Ser &  27.7 &  +45.7 &  11.1 & 28.85 &  18.20 & $-$1.34 $^{+0.24}_{-0.58}$ \\
WD1821$-$131 & 20 &  18.1 & $-$0.0 &  $-$1.816        & $\gamma$ Oph &  28.0 &  +15.4 &  29.1 & 18.25 &  18.46 & $-$1.50 $^{+0.32}_{-0.67}$ \\
           &    &        &        & & HR 6748 &  356.0 &  $-$7.3 &  17.4 & 23.25 &  18.13 & $-$1.60 $^{+0.02}_{-0.02}$ \\
WD1826$-$045 & 25 & 26.4 & +2.9  &    $-$1.583         & 70 Oph  &  29.9 &  +11.4 &   5.1 &  9.20 &  17.99 & $-$1.21 $^{+0.08}_{-0.09}$ \\
           &    &        &        & & $\gamma$ Oph &  28.0 &  +15.4 &  29.1 & 12.61 &  18.46 & $-$1.50 $^{+0.32}_{-0.67}$\\
           &    &        &        & & $\zeta$ Aql  &  46.9 &   +3.3 &  25.5 & 20.49 &  18.18 & $-$1.71 $^{+0.32}_{-0.68}$\\
WD1858+393 & 35 &  69.8 & +15.2  &  $-$3.873:          & SAO 68491 &   67.0 &   +6.5 &  32.0 &  9.14 &  18.88 & $-$1.12 $^{+0.39}_{-1.09}$\\
           &    &        &        & & $\delta$ Cyg &  78.7 &  +10.2 &  52.4 & 10.07 &  18.59 & $-$1.62 $^{+0.32}_{-0.67}$\\
WD2115$-$560 & 22 & 340.5 & $-$42.7 &  +0.087:         & $\alpha$ Gru & 350.0 & $-$52.5 &  31.1 & 11.66 &  18.72 & $-$1.26 $^{+0.24}_{-0.60}$\\
           &    &        &        & & SAO 254993 & 324.9 & $-$38.9 &  20.5 & 12.41 &  18.82 & $-$0.98 $^{+0.02}_{-0.02}$\\
HS2132+0941 & 79 & 63.8 & $-$29.6 &   $-$0.935          & $\alpha$ Del &  60.3 & $-$15.3 &  73.8 & 14.65 &  18.31 & $-$2.04 $^{+0.32}_{-0.67}$\\
           &    &        &        & & $\theta$ Aql &  41.6 & $-$18.1 &  88.0 & 23.28 &  18.57 & $-$1.86 $^{+0.32}_{-0.67}$\\
WD2149+021 & 25 &  60.1 & $-$37.8 &   $-$4.467:          & $\theta$ Peg &  67.4 & $-$38.7 &  29.6 &  5.83 &  18.12 & $-$1.84 $^{+0.32}_{-0.69}$\\
           &    &        &        & & V376 Peg &   76.8 & $-$28.5 &  47.1 & 16.76 &  18.37 & $-$1.79 $^{+0.03}_{-0.03}$\\
HE2221$-$1630 &  52 & 42.9 & $-$54.2 &  $-$0.714         & $\iota$ Cap &   33.6 & $-$40.8 &  66.1 & 14.72 &  18.63 & $-$1.68 $^{+0.17}_{-0.24}$\\
           &    &        &        & & $\gamma$ Aqr &  62.2 & $-$45.8 &  48.4 & 14.90 &  17.60 & $-$2.58 $^{+0.32}_{-0.69}$\\
HS2229+2335 & 91 & 86.1 & $-$28.8 &  +0.222         & $\zeta$ Peg &   78.9 & $-$40.7 &  64.1 & 13.25 &  18.05 & $-$2.25 $^{+0.32}_{-0.68}$\\
           &    &        &        & & $\alpha$ Del &  60.3 & $-$15.3 &  73.8 & 27.40 &  18.31 & $-$2.04 $^{+0.32}_{-0.67}$\\
WD2326+049 & 19 &  88.2 & $-$52.0 &  +0.340         & HR 8 &  111.3 & $-$32.8 &  13.7 & 25.45 &  18.29 & $-$1.34 $^{+0.01}_{-0.01}$\\
           &    &        &        & & PW And &  114.6 & $-$31.4 &  21.9 & 28.20 & 18.05 & $-$1.78 $^{+0.09}_{-0.11}$\\
\enddata
\tablecomments{$\log n$(H)$_{\rm pred}$ is the theoretically predicted density required to explain the observed metals in the atmosphere of the white dwarf.
WD0235+064, WD0846+346, WD1407+425, WD1858+393, WD2115-560, and WD2149+021 lack space velocity measurements, and their $\log n$(H)$_{\rm pred}$ estimates
are inaccurate, and therefore are not shown in subsequent figures. }
\end{deluxetable}
\clearpage
\begin{figure}
\plotone{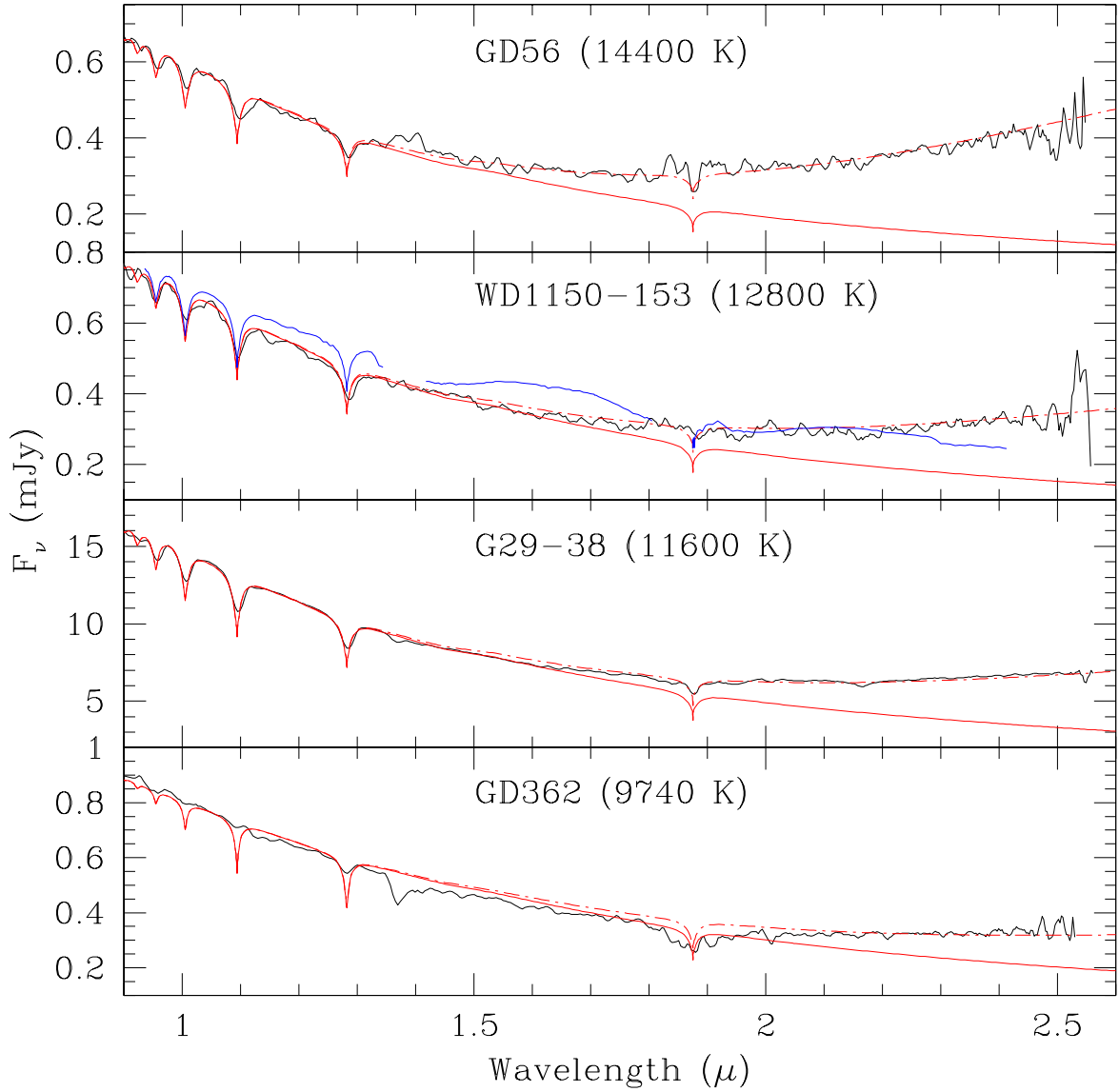}		
\caption{The IRTF spectra of white dwarfs with circumstellar debris disks (black lines). The expected
near-infrared fluxes from each star (red solid line) and composite white dwarf + 900 K dust templates (dashed-dotted line) are shown
in each panel. The second panel also shows a composite white dwarf + L6 dwarf template for WD1150-153 (blue line).}
\end{figure}

\clearpage
\begin{figure}
\plotone{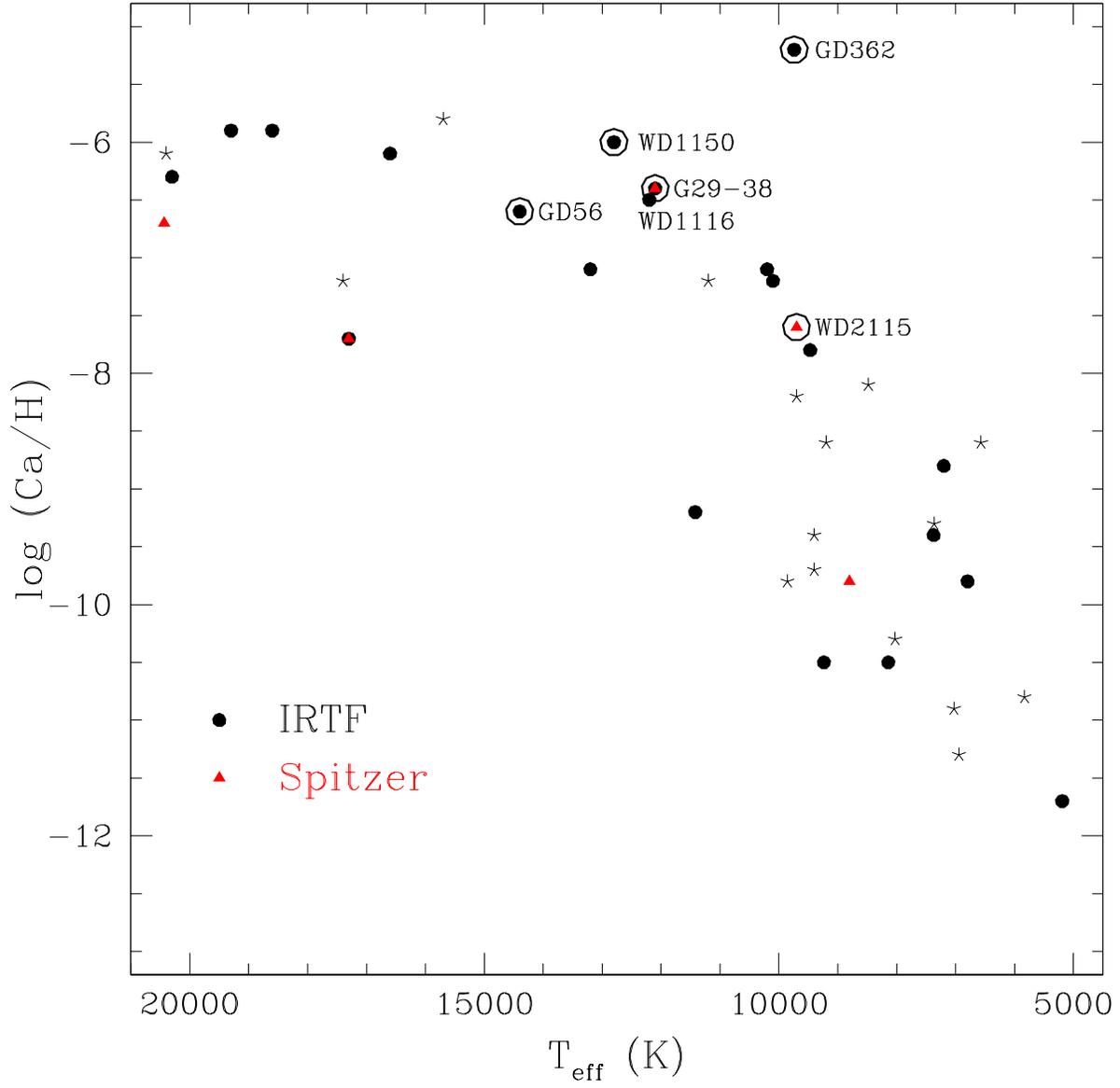}		
\caption{Calcium abundances versus effective temperatures for the objects observed at the IRTF (filled circles; Kilic et al. 2006) and
Spitzer/IRAC (filled triangles; from Mullally et al. 2006). The rest of the DAZs
from KW06 are shown with star symbols. White dwarfs with circumstellar debris disks are
marked with open circles.}
\end{figure}

\clearpage
\begin{figure}
\plotone{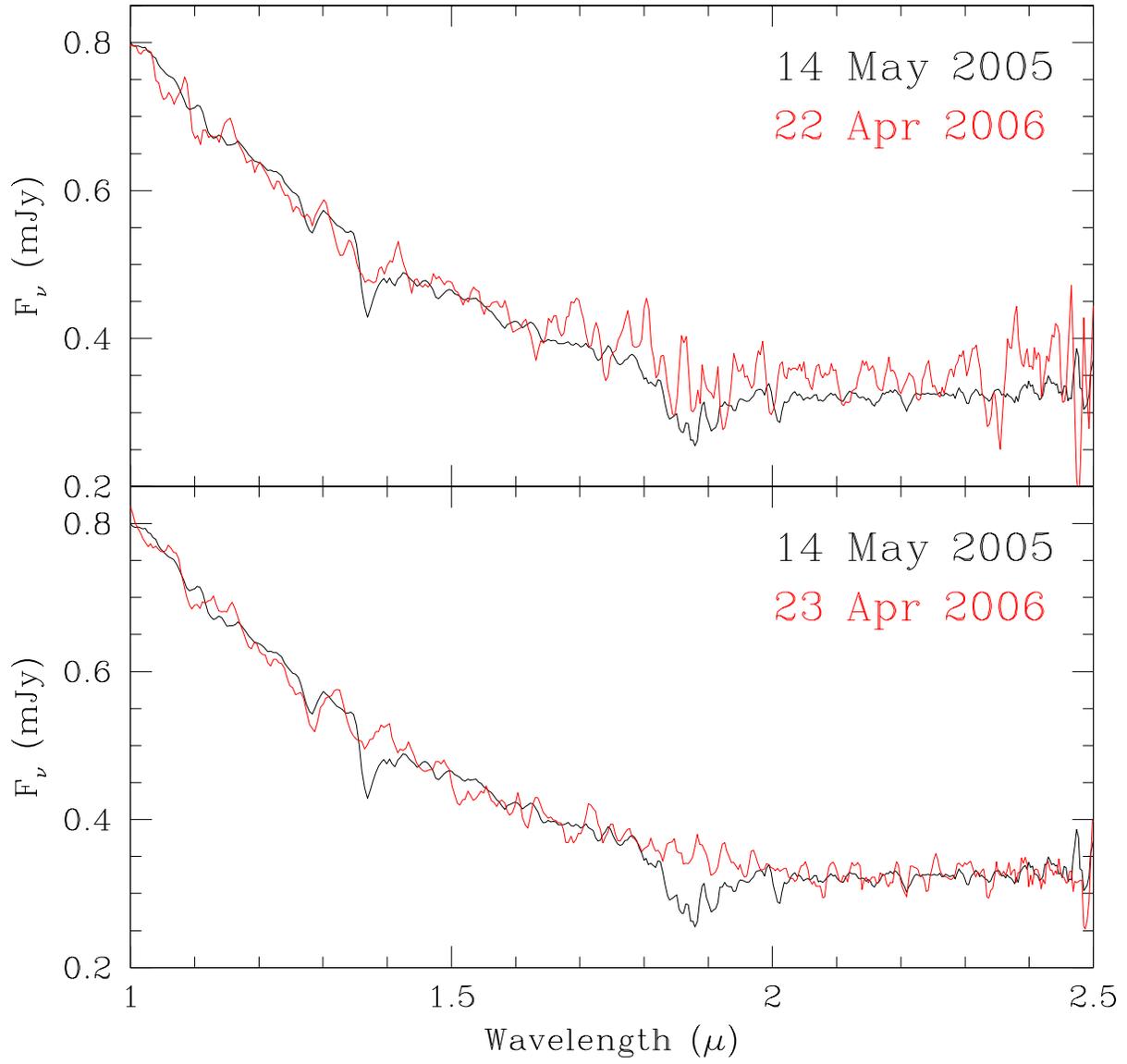}		
\caption{Near infrared spectra of GD362 in May 2005 and April 2006.}
\end{figure}

\clearpage
\begin{figure}
\plotone{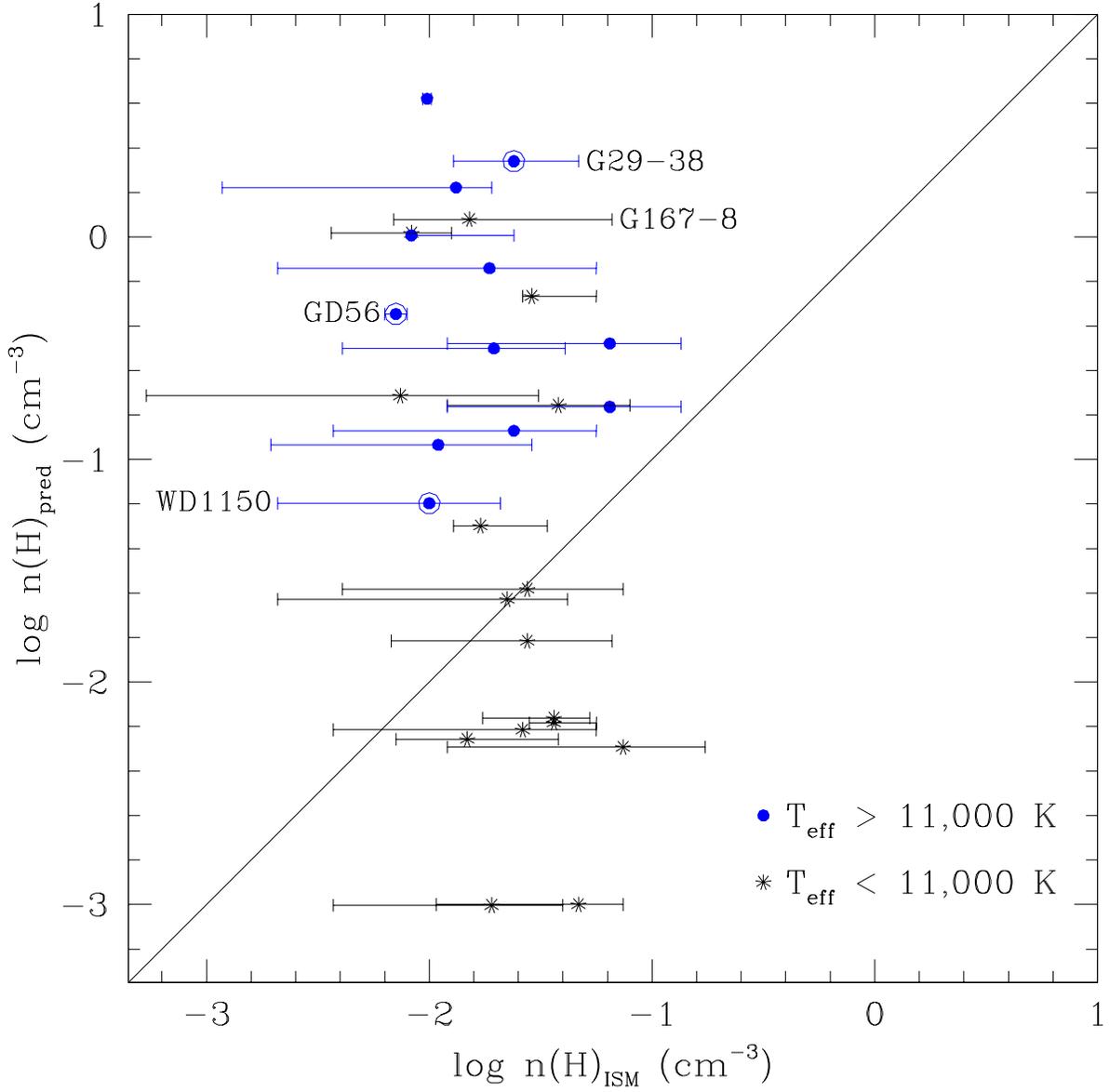}		
\caption{The theoretically predicted versus observed ISM densities
around known DAZ white dwarfs. The required ISM densities to explain
metals in DAZ white dwarfs assume the validity of the Bondi-Hoyle
accretion formula (KW06). White dwarfs with $T_{\rm eff}>$ 11,000 K
and $T_{\rm eff}<$ 11,000 K are shown as filled circles and star
symbols, respectively. The ``error bars'' include
the propagated errors of column density, distance, and abundance ratio
to hydrogen and they show the observed range of ISM volume densities around each
white dwarf. White dwarfs with circumstellar debris disks are marked
with open circles. Farihi et al.'s debris disk candidate (G167-8) is
also labeled.}
\end{figure}

\clearpage
\begin{figure}
\plotone{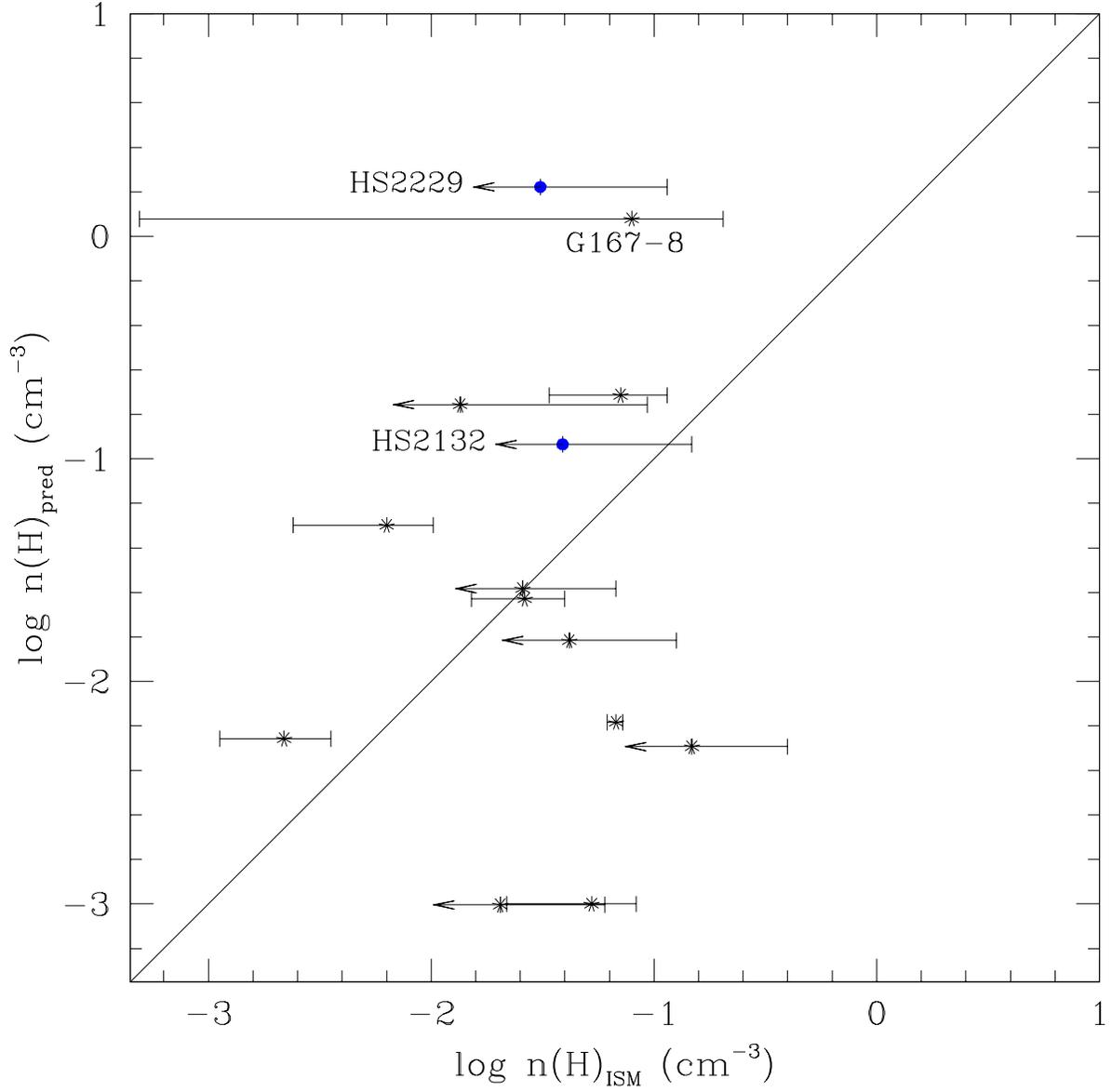}		
\caption{Same as Figure 4, except the ISM densities are measured in the vicinity of the white dwarfs.}
\end{figure}

\end{document}